\documentclass[aps,prl,twocolumn,nofootinbib,tightenlines,showpacs,%
showkeys,floatfix,tightenlines,preprintnumbers]{revtex4}

\usepackage{epsfig,bm}

\begin{document}



\title{Decay modes of ideally mixed narrow pentaquark states}

\author{Su Houng Lee}%
\email{suhoung@phya.yonsei.ac.kr}

\author{Hungchong Kim}%
\email{hung@phya.yonsei.ac.kr}
\altaffiliation[Present Address: ]{Department of Physics,
Pohang University of Science and Technology, Pohang, 790-784, Korea}

\author{Yongseok Oh}%
\email{yoh@physast.uga.edu}
\altaffiliation[Present Address: ]{Department of Physics and Astronomy,
University of Georgia, Athens, GA 30602, U.S.A.}

\affiliation{Institute of Physics and Applied Physics,
Yonsei University, Seoul 120-749, Korea}


\begin{abstract}
We investigate the decay modes of pentaquark baryons for both the
antidecuplet and the octet states, which were recently claimed by
Close and Dudek to be potentially narrow due to a hidden selection
rule.
We discuss how ideal mixing between the pentaquark octet and
antidecuplet states can be related to the OZI rule, which also
naturally leads to certain selection rules in the pentaquark
decays.
We then introduce a tensor representation for the pentaquark states, and
present the possible decay modes for both the unmixed and ideally mixed
pentaquark octet and antidecuplet states.
The exclusive decay modes can be used to experimentally search for the
other pentaquark states.
\end{abstract}


\maketitle


\section{Introduction}

The recent discovery of the $\Theta^+$ baryon by LEPS Collaboration at
SPring-8 \cite{LEPS03}, which has been subsequently confirmed by several
groups \cite{CLAS03-b,CLAS03-d,SAPHIR03,DIANA03,ADK03,HERMES03a,SVD04},
initiated a lot of theoretical works in the field of hadron physics.
Experimentally, $\Theta^+$ is observed to have a mass of 1540 MeV and a
decay width of $< 25$ MeV, but there are indications that the width should
in fact be much smaller \cite{Nuss03,ASW03-ASW03b,HK03,CT04}.
The $\Theta^+(1540)$ together with the recently discovered $\Xi^*(1862)$
\cite{NA49-03} (See also Ref.~\cite{FW04}.) are expected to be the members
of the antidecuplet pentaquark states \cite{DPP97}.
At this stage, a pressing issue is to identify the other members of the
antidecuplet and the pentaquark octet states.
What is also potentially interesting about these pentaquark states is
their small decay widths.
Although $\Theta^+$ can decay into a kaon and a nucleon, its width is
smaller than that obtained from any reasonable estimate based on the
assumption that it is a meson-nucleon quasi-bound state \cite{Capstick}.
Hence, the picture of pentaquark states by Jaffe and Wilczek \cite{JW03}
based on a strong diquark states seems favorable, as the regrouping of the
color, flavor-spin and spatial wavefunctions of the diquarks and an
antiquark into the nucleon and kaon is expected to be quite
costly \cite{JM03,KL03a,CCKN03c,CD04,MSS04}.
In this picture, one can make two further interesting assumptions.
The first is the fall-apart mechanism in their decays \cite{CD04} and the
other is the ideal mixing \cite{JW03}.
As we will discuss, both assumptions are the consequences of a generalized
OZI rule.

In the formalism of Jaffe and Wilczek \cite{JW03}, two quarks form
a color triplet boson state, two of which with a relative $p$-wave
form a color $\overline{\bf 3}$ state.
These two diquarks then combine with an antiquark to form a color singlet
state.
If $\Theta^+$ belongs to antidecuplet, then it is an isospin singlet
state.
In this case, the diquarks, each in flavor $\overline{\bf 3}$, will
form a $\overline{\bf 6}$, which then combines with an antiquark
to give a $\overline {\bf 10}$.
However, since $\overline{\bf 6} \otimes \overline {\bf 3} =
\overline {\bf 10} \oplus {\bf 8}$, we also expect a pentaquark octet
naturally, which differs from the soliton model picture \cite{DPP97}.
As has been discussed by Close and Dudek in Ref.~\cite{CD04}, the
pentaquark octet is also expected to share a common feature with the
antidecuplet and have a small decay width.
The decays of these pentaquark octet and antidecuplet states are expected
to proceed through a ``fall-apart" mechanism.
This mechanism states that there is no annihilation or creation of quark
pairs and the decay proceeds through a recombination of quarks from the
strongly correlated diquarks into the final decay products.
This follows from a generalized OZI rule in pentaquark decays, which also
leads to ideal mixing between the octet and antidecuplet pentaquark states.
If we apply this rule to the decay for both the pentaquark octet and
antidecuplet decays, we obtain the ``selection rules" advocated by Close
and Dudek in Ref.~\cite{CD04}.
As we will show below, one can generalize the rule by using the tensor
representation for the pentaquark states.

In this paper, we start with how the generalized OZI rule leads to ideal
mixing of the pentaquark octet and antidecuplet states, and to the
fall-apart mechanism for the pentaquark decays.
We then introduce a tensor formalism for the pentaquarks and derive the
selection rules for their decays.
Using the tensor method, we present all the possible decay modes of the
pentaquark octet and antidecuplet states into the baryon octet and meson
octet.
We also discuss why certain decays are not allowed.
In addition, the pentaquark octet decays into baryon decuplet and
meson octet are shown to be prohibited by the OZI rule even if
they are allowed energetically and by SU(3) symmetry.
The SU(3) symmetry breaking effects of this decay is also discussed.

\section{Ideal mixing}

\subsection{Quantum mechanical example}

The selection rule follows from the fall-apart mechanism, which effectively
is the OZI rule for the pentaquark decays.
Here, we investigate the idea of the generalized OZI rule in the present
context and show that it also naturally leads to ideal mixing among the
pentaquark states.

Let us start with a simple pedagogical example of two-level quantum
mechanical system, with the Hamiltonian
\begin{eqnarray}
H=\left(
\begin{array}{cc}
E_8 & \Delta \\
\Delta & E_1
\end{array}\right).
\label{hamiltonian}
\end{eqnarray}
Here, we call the two states as
\begin{eqnarray}
\psi_8= \left(
\begin{array}{c}
1 \\
0
\end{array}\right), \qquad
\psi_1= \left(
\begin{array}{c}
0 \\
1
\end{array}\right) .
\end{eqnarray}

If there is a mixing in the mass matrix as in Eq.~(\ref{hamiltonian}), the
physical states are the eigenstates of the Hamiltonian.
The two eigenvalues are
\begin{eqnarray}
\lambda_\pm=\frac{1}{2} \bigg( E_8+E_1\pm \sqrt{(E_8-E_1)^2+4
\Delta^2} \bigg),
\end{eqnarray}
and the corresponding eigenvectors are
\begin{eqnarray}
\phi & = & \cos \theta \, \psi_8 + \sin \theta \, \psi_1, \nonumber \\
\omega & = & -\sin \theta \, \psi_8 + \cos \theta \, \psi_1,
\label{mixing}
\end{eqnarray}
where
\begin{eqnarray}
\tan \theta = {\lambda_+-E_8 \over \Delta},  \label{angle1}
\end{eqnarray}
or equivalently,
\begin{eqnarray}
E_8-E_1  =  {1- \tan^2 \theta \over \tan \theta} \Delta.
\label{angle2}
\end{eqnarray}
Formally, one can also write $E_8=\langle \psi_8|H|\psi_8 \rangle,
E_1=\langle \psi_1|H|\psi_1 \rangle, \Delta=\langle \psi_1|H|\psi_8 \rangle$.
An alternative way to obtain the mixing angle, Eq.~(\ref{angle1}) or
Eq.~(\ref{angle2}), is substituting Eq.~(\ref{mixing}) into the following
condition,
\begin{eqnarray}
\langle \phi |H | \omega \rangle=0.
\label{condition}
\end{eqnarray}

In the case of ideal mixing, the mixing angle thus obtained also
isolates the strange and nonstrange quark-antiquark pair into
separate wavefunctions.
That is, in such an ideal case, one finds that $\phi$ is a pure
$\bar{s}s$ state while $\omega$ is purely a nonstrange $\bar{q}q$ state.
Stated in a diagrammatic language, ideal mixing follows when connected
quark graphs dominate over disconnected quark graphs, which is also known as
the OZI rule.
This is so because the eigenvalues satisfy $\lambda_\pm \gg \langle \bar{s}s
|H| \bar{q}q \rangle =0$, where $\lambda_+ = \langle \bar{s}s |H|\bar{s}s
\rangle$ and $ \lambda_- = \langle \bar{q}q |H|\bar{q}q \rangle$.

\subsection{Vector and pseudoscalar channel}

First, one should note that a prerequisite for mixing is SU(3)
symmetry breaking.
That is, the Hamiltonian in Eq.~(\ref{hamiltonian}) should have explicit
symmetry breaking such that
\begin{eqnarray}
\Delta = \langle \psi_1|H|\psi_8 \rangle \neq  0 .
\label{delta}
\end{eqnarray}

For the vector meson channel, the physical wavefunctions obtained
from isolating the strange and nonstrange quark-antiquark parts
are,
\begin{eqnarray}
\phi & =& -{\bar s} s =  \sqrt{\frac{2}{3}} \psi_8 -
\sqrt{\frac{1}{3}} \psi_1,
\nonumber \\
\omega & = & \frac{1}{\sqrt{2}}({\bar u}u+{\bar d}d)=
\sqrt{\frac{2}{3}} \psi_1+ \sqrt{\frac{1}{3}} \psi_8,
\end{eqnarray}
where
\begin{eqnarray}
\psi_1 & =& \frac{1}{\sqrt{3}}({\bar u} u+{\bar d}d + {\bar s} s),
\nonumber \\
\psi_8 & = &  \frac{1}{\sqrt{6}}({\bar u} u+{\bar d}d -2 {\bar s}
s).
\end{eqnarray}
Hence, the mixing angle coming from isolating the nonstrange and
strange part in the wavefunction is $\tan \theta = -1/\sqrt{2}$,
which is also close to the  physical mixing angle.
Hence, ideal mixing in the vector channel implies
\begin{eqnarray}
\langle \bar{s}\gamma_\mu s |H| \bar{u} \gamma_\mu
u+\bar{d}\gamma_\mu d \rangle =0,
\label{condition1-1}
\end{eqnarray}
where we have included $\gamma_\mu$ to represent that the states are
vector mesons.

For the pseudoscalar channel, the situation is quite different and the
mixing angle is very small, suggesting that the
disconnected quark lines are not suppressed,
\begin{eqnarray}
\langle \bar{s} \gamma_5 s |H| \bar{u} \gamma_5 u +
\bar{d} \gamma_5 d \rangle \neq \mbox{small}.
\label{condition1-2}
\end{eqnarray}
The reason for such a large difference between Eqs.~(\ref{condition1-1})
and (\ref{condition1-2}) are relatively well understood in both the
perturbative and nonperturbative regime.
Perturbatively, three gluon intermediate state are needed to connect the
vector currents in Eq.~(\ref{condition1-1}), while only two gluon states
are needed in Eq.~(\ref{condition1-2}).
In the nonperturbative regime, at least two classical instanton
configuration is needed to connect the vector currents, while only a
single instanton is needed in the pseudoscalar channel \cite{Shur93}.

\subsection{Pentaquarks}

The realization of ideal mixing between pentaquark octet and antidecuplet
depends on whether a condition similar to Eq.~(\ref{condition}) is
fulfilled by the pentaquark currents.
We first note that the wavefunctions of the nucleon with positive charge
in pentaquark antidecuplet and octet are obtained as
\begin{eqnarray}
N_{\overline{10}}^+ & =& \frac{1}{\sqrt{3}}\bigg(
[ud][ud]\bar{d}+\sqrt{2} [ud][us]_+ \bar{s} \bigg), \nonumber \\
N_8^+ & =& \frac{1}{\sqrt{3}}\bigg( -\sqrt{2} [ud][ud] \bar{d}
+[ud][us]_+ \bar{s} \bigg).
\label{wave0}
\end{eqnarray}
Here, $[ud]=\frac{1}{\sqrt{2}}( ud-du)$ and
$[A][B]_+=\frac{1}{\sqrt{2}}([A][B]+[B][A] )$.
We also have suppressed color and Dirac indices.
As before, the ideal mixing is achieved when the following
rule is satisfied:
\begin{eqnarray}
\langle [ud][ud]\bar{d}| H |[ud][us]_+ \bar{s} \rangle = {\rm
small}.
\label{cond-penta}
\end{eqnarray}
We call it ``generalized OZI rule", as none of the previous theoretical
arguments, which justify the OZI rule in the meson sector, applies in
this case.
It is through experiments that we can confirm ideal mixing.
But since ideal mixing is a consequence of the ``generalized OZI rule" it is
important to probe other consequences of this rule.

A useful result, that can be directly verified experimentally, is the
characteristic features of decay modes of the pentaquark baryons due to
the ``generalized OZI rule".
Consider the decays of the pentaquark octet and antidecuplet into a kaon
and a hyperon.
The respective coupling constants are proportional to the overlap of the
hadron wavefunctions as
\begin{eqnarray}
g_8^{} & \propto & \langle N_8 | K \Sigma \rangle,
\nonumber \\
g_{\overline{10}}^{} & \propto & \langle N_{\overline{10}} | K
\Sigma \rangle.
\end{eqnarray}
Assuming the ``generalized OZI rule", the decays are dominated by
the pentaquark wavefunction in Eq.~(\ref{wave0}) that contains
$\bar{s}s$, which have common space-spin wave function \cite{Cohen04}.
Hence the ratio between the respective matrix elements are determined just
by SU(3) factors as
\begin{eqnarray}
{\langle N_8 | K \Sigma \rangle \over \langle N_{\overline{10}} |
K \Sigma \rangle
  }  & = & -\sqrt{2},
\nonumber \\
{\langle N_8 | \pi N \rangle \over \langle N_{\overline{10}} | \pi
N \rangle   }  & = & \frac{1}{\sqrt{2}}.
\label{fractions}
\end{eqnarray}
Such rules are equivalent to the fall apart mechanism introduced
by Close and Dudek \cite{CD04} and are a consequence of the
``generalized OZI rule", which also predicts ideal mixing among
the pentaquark octet and antidecuplet.

\section{Tensor notation}

In this section, we introduce the tensor notation for the pentaquark
states and derive selection rules for their decays assuming the OZI
rule.
The pentaquark antidecuplet and octet states can be represented
by $T^{ijk}$ and $P_i^j$, respectively, where $i,j,k$ are the flavor
indices $u,d,s$.
We can obtain the SU(3) relation in their decays into either baryon
decuplet $D_{ijk}$ or normal baryon octet $B_i^j$ made of three quarks
and a meson octet $M_i^j$ by constructing SU(3) invariant Lagrangians.
For example, the SU(3) invariant relations for the decay of pentaquark
antidecuplet into a normal baryon octet and meson octet can be obtained
as \cite{OKL03b}%
\footnote{Usually, a factor $-i$ is multiplied to the interaction
Lagrangian.
Since it does not affect our discussion, we omit this factor.
The Lorentz structure is also suppressed.}
\begin{eqnarray}
\mathcal{L}_{\overline{10}}= -g_{\overline{10}}^{} \epsilon^{ilm}
\overline{T}_{ijk} B_l^j M_m^k + \mbox{(H.c.)}, \label{int-1}
\end{eqnarray}
where the particle identification of $T^{ijk}$ can be found in
Ref.~\cite{OKL03b} and the forms for $B_i^j$ and $M_i^j$ can be
found, e.g., in Ref.~\cite{SR99}.
By expanding Eq.~(\ref{int-1}), we explicitly have
\begin{eqnarray}
\mathcal{L}_{\overline{10}} &=&
-\sqrt6 \, \overline{\Theta} \, \overline{K}_c N
- \overline{N}_{\overline{10}}^{} \bm{\tau}\cdot \bm{\pi} N
\nonumber \\ && \mbox{}
+ \sqrt3 \, \eta_8^{} \overline{N}_{\overline{10}}^{} N
- \sqrt3 \, \overline{N}_{\overline{10}}^{} K \Lambda
\nonumber \\ && \mbox{}
+ \overline{N}_{\overline{10}}^{} \bm{\tau} \cdot \bm{\Sigma} K
+ i \left( \overline{\bm{\Sigma}}_{\overline{10}} \times \bm{\Sigma} \right)
\cdot \bm{\pi}
\nonumber \\ && \mbox{}
- \sqrt3 \, \overline{\bm{\Sigma}}_{\overline{10}} \cdot \bm{\pi} \Lambda
+ \sqrt3 \, \eta_8^{} \overline{\bm{\Sigma}}_{\overline{10}} \cdot \bm{\Sigma}
\nonumber \\ && \mbox{}
- \overline{K}_c \overline{\bm{\Sigma}}_{\overline{10}} \cdot \bm{\tau} \, \Xi
- \overline{K}\, \overline{\bm{\Sigma}}_{\overline{10}} \cdot \bm{\tau} N
\nonumber \\ && \mbox{}
- \sqrt3 \, \overline{\Xi}_{\overline{10}} \bm{T} \cdot \bm{\pi}\, \Xi
+ \sqrt3 \, \overline{\Xi}_{\overline{10}} \bm{T} \cdot \bm{\Sigma} K_c
\nonumber \\ &&
\mbox{} + \mbox{(H.c.)},
\end{eqnarray}
where the overall factor $g_{\overline{10}}^{}$ is understood and
\begin{eqnarray}
&& N = \left( \begin{array}{c} p \\ n \end{array} \right), \quad
\Xi = \left( \begin{array}{c} \Xi^0 \\ \Xi^- \end{array} \right),
\quad
K = \left( \begin{array}{c} K^+ \\ K^0 \end{array} \right),
\nonumber \\ &&
K_c = \left( \begin{array}{c} \bar{K}^0 \\ -K^- \end{array} \right),
\quad
\bm{\tau} \cdot \bm{\pi} = \left(
\begin{array}{cc} \pi^0 & \sqrt2 \pi^+ \\ \sqrt2 \pi^- & -\pi^0
\end{array} \right),
\nonumber \\
\end{eqnarray}
etc. The isospin transition operator $\bm{T}$ is defined as
\cite{CN69}
\begin{eqnarray}
&& T^{(+1)} = \frac{1}{\sqrt3}
\left( \begin{array}{cc} \sqrt6 & 0 \\ 0 & \sqrt2 \\ 0 & 0 \\ 0 & 0
\end{array} \right), \quad
T^{0} = \frac{1}{\sqrt3} \left( \begin{array}{cc} 0 & 0 \\ 2 & 0 \\ 0 & 2
\\ 0 & 0 \end{array} \right),
\nonumber \\ &&
T^{(-1)} =
\frac{1}{\sqrt3} \left( \begin{array}{cc} 0 & 0 \\ 0 & 0 \\ \sqrt2
& 0 \\ 0 & \sqrt6 \end{array} \right),
\end{eqnarray}
so that $\bm{T} \cdot \bm{\pi} = -T^{(+1)} \pi^+ + T^{(-1)} \pi^-
+ T^0 \pi^0$ and so on.

\begin{table}[t]
\centering
\begin{ruledtabular}
\begin{tabular}{cc|cc|cc|cc}
\multicolumn{2}{c}{$\Theta^+$} & \multicolumn{2}{c}{$N_{\overline{10}}^+$} &
\multicolumn{2}{c}{$N_{\overline{10}}^0$} &
\multicolumn{2}{c}{$\Sigma_{\overline{10}}^+$}
\\ \hline
$K^+ n$ & $\sqrt{6}$ & $\pi^+ n$ & $-\sqrt2$ & $\pi^0 n$ & $1$ &
$\pi^+ \Lambda$ & $-\sqrt{3}$ \\
$K^0 p$ & $-\sqrt6$ & $\pi^0 p$ & $-1$ & $\pi^- p$ & $-\sqrt2$
& $\pi^+ \Sigma^0$ & $1$ \\
& & $\eta_8^{} p$ & $\sqrt{3}$ & $\eta_8^{} n$ & $\sqrt{3}$ &
$\pi^0 \Sigma^+$ & $-1$ \\
& & $K^+ \Lambda$ & $-\sqrt{3}$ & $K^+ \Sigma^-$ & $\sqrt2$ &
$\eta_8^{}
\Sigma^+$ & $\sqrt{3}$ \\
& & $K^+ \Sigma^0$ & $1$ & $K^0 \Lambda$ &
$-\sqrt{3}$ & $K^+ \Xi^0$ & $\sqrt{2}$ \\
& & $K^0 \Sigma^+$ & $\sqrt{2}$ & $K^0 \Sigma^0$ & $-1$ &
$\bar{K}^0 p $ & $-\sqrt{2}$ \\ \hline
\multicolumn{2}{c}{$\Sigma_{\overline{10}}^0$} &
\multicolumn{2}{c}{$\Sigma_{\overline{10}}^-$} &
\multicolumn{2}{c}{$\Xi^+_{\overline{10}}$} &
\multicolumn{2}{c}{$\Xi^0_{\overline{10}}$} \\ \hline $\pi^+ \Sigma^-$ & $-1$
& $\pi^0 \Sigma^-$ &
$1$ & $\pi^+\Xi^0$ & $\sqrt{6}$ & $\pi^+ \Xi^-$ & $\sqrt2$ \\
$\pi^0 \Lambda$ & $-\sqrt{3}$ & $\pi^- \Lambda$ & $-\sqrt{3}$
& $\bar{K}^0 \Sigma^+$ & $-\sqrt6$ & $\pi^0 \Xi^0$ & $-2$ \\
$\pi^- \Sigma^+$ & $1$ & $\pi^- \Sigma^0$ &
$-1$ & & & $\bar{K}^0 \Sigma^0$ & $2$ \\
$\eta_8^{}\Sigma^0$ & $\sqrt{3}$ & $\eta_8^{} \Sigma^-$ &
$\sqrt{3}$ &
& & $K^-\Sigma^+$ & $\sqrt2$ \\
$K^+ \Xi^-$ & $-1$ & $K^0 \Xi^-$ & $-\sqrt2$ & & & & \\
${K}^0 \Xi^0$ & $-1$ & $K^- n$ & $-\sqrt2$ & & & & \\
$\bar{K}^0 n$ & $1$ & & & & & & \\
${K}^- p$ & $-1$ & & & & & & \\ \hline
\multicolumn{2}{c}{$\Xi^-_{\overline{10}}$} &
\multicolumn{2}{c}{$\Xi^{--}_{\overline{10}}$} & \multicolumn{2}{c}{} &
\multicolumn{2}{c}{} \\ \hline
$\pi^0 \Xi^-$ & $-2$ & $\pi^- \Xi^-$ & $-\sqrt6$ & & & & \\
$\pi^- \Xi^0$ & $-\sqrt2$ & $K^- \Sigma^-$ & $-\sqrt6$ & & & & \\
$\bar{K}^0 \Sigma^-$ & $\sqrt2$ & & & & & & \\
${K}^- \Sigma^0$ & $-2$ & & & & & & \\
\end{tabular}
\end{ruledtabular}
\caption{Couplings of the pentaquark antidecuplet with the baryon
octet and pseudoscalar meson octet.
Multiplying the universal coupling constant $g_{\overline{10}}^{}$ is
understood.}
\label{table-10bar}
\end{table}

The obtained couplings are listed in Table~\ref{table-10bar}.
Moreover, as shown in Ref.~\cite{OKL03b}, it is straightforward to
show that one can not form a coupling between the antidecuplet and
baryon decuplet and meson octet, which suggests that the $N(1710)$
can not be a member of the pure antidecuplet as it has large decay
width into the $\Delta\pi$ channel \cite{OKL03b,JW03,CD04}.

In order to construct SU(3) invariant interactions for the pentaquark
octet, which respects the OZI rule, we have to be careful with the
contraction of the SU(3) tensors.
To do that, we note that the pentaquark octet and antidecuplet come from
the $\overline{\bf 6}$ of two diquarks and $\overline{\bf 3}$ of one
antiquark, i.e., $\overline {\bf 6} \otimes \overline {\bf 3} =
\overline {\bf 10} \oplus {\bf 8}$.
This follows from
\begin{eqnarray}
S^{ij} \otimes \bar{q}^k=T^{ijk}\oplus S^{[ij,k]},
\label{tensor1}
\end{eqnarray}
where $ijk$ are symmetric in $T^{ijk}$ (antidecuplet), and symmetric in
$ij$ but antisymmetric with $k$ in $S^{[ij,k]}$.
Obviously, the last part, being an octet representation, can be
replaced by two-index field $P^j_i$ such as
\begin{eqnarray}
S^{[ij,k]}=\epsilon^{ljk} P_l^i+\epsilon^{lik}P_l^j.
\label{tensor2}
\end{eqnarray}

When we construct an effective interaction of the pentaquark octet
$P$ with the usual baryon octet $B$ and meson octet $M$, we have
two couplings, namely, $f$ type and $d$ type,
\begin{eqnarray}
\mathcal{L}_{8} &=& g_8^{} (d-f) \overline{P}_i^l B_l^k M_k^i +
g_8^{}(d+f) \overline{P}_i^l M_l^k B_k^i
\nonumber \\ && \mbox{}
+ \mbox{(H.c.)} .
\label{fd}
\end{eqnarray}
However, as has been noted by Close and Dudek \cite{CD04}, such
form does not represent the fall-apart mechanism.
To include that mechanism, one has to note that the index $k$ in
Eq.~(\ref{tensor2}), the index for the antiquark, should be contracted
with the antiquark index of the meson field to represent the fall-apart
mechanism, as the usual baryon $B$ does not contain an antiquark.
Hence, the interaction should follow the same form as in Eq.~(\ref{int-1}),
\begin{eqnarray}
\mathcal{L}_{8}= g_{8}^{} \epsilon^{ilm} \overline{S}_{[ij,k]}
B^j_l M_m^k + \mbox{(H.c.)}.
\label{interaction2}
\end{eqnarray}
Substituting Eq.~(\ref{tensor2}) into Eq.~(\ref{interaction2}), one has
\begin{eqnarray}
\mathcal{L}_{8} = 2 g_8^{} \overline{P}^m_i B^i_l M^l_m + g_8
\overline{P}^m_i M^i_l B^l_m+ \mbox{(H.c.)},
\end{eqnarray}
which is equivalent to Eq.~(\ref{fd}) with $f=1/2$ and $d=3/2$.
Therefore, the OZI rule makes a special choice on the $f/d$ ratio as
$f/d=1/3$ \cite{CD04}.
With this information, all the possible interactions of the pentaquark
octet baryons can be calculated as
\begin{eqnarray}
\mathcal{L}_{8} &=&
g_{N_8^{}N\pi}^{} \overline{N}_8^{} \bm{\tau} \cdot \bm{\pi} N
+ g_{N_8^{}N\eta_8^{}}^{} \eta_8^{} \overline{N}_8^{} N
\nonumber \\ && \mbox{}
+ g_{N_8^{}\Lambda K}^{} \overline{N}_8^{} K \Lambda
+ g_{N_8^{}\Sigma K}^{} \overline{N}_8^{} \bm{\tau}\cdot \bm{\Sigma} K
\nonumber \\ && \mbox{}
+ i g_{\Sigma_8^{} \Sigma \pi}^{} \left( \overline{\bm \Sigma}_8^{} \times
\bm{\Sigma} \right) \cdot \bm{\pi}
+ g_{\Sigma_8^{} \Sigma \eta_8^{}}^{} \eta_8 \overline{\bm{\Sigma}}_8 \cdot
\bm{\Sigma}
\nonumber \\ && \mbox{}
+ g_{\Sigma_8^{} \Lambda \pi}^{} \overline{\bm{\Sigma}}_8 \cdot \bm{\pi}
\Lambda
+ g_{\Sigma_8^{} \Xi K}^{} \overline{K}_c \overline{\bm{\Sigma}}_8 \cdot
\bm{\tau} \,\Xi
\nonumber \\ && \mbox{}
+ g_{\Sigma_8^{} NK}^{} \overline{K} \, \overline{\bm{\Sigma}}_8 \cdot
\bm{\tau} N
+ g_{\Lambda_8^{} \Sigma \pi}^{} \overline{\Lambda}_8 \bm{\Sigma} \cdot
\bm{\pi}
\nonumber \\ && \mbox{}
+ g_{\Lambda_8^{} \Lambda \eta_8^{}}^{} \eta_8 \overline{\Lambda}_8 \Lambda
+ g_{\Lambda_8^{} \Xi K}^{} \overline{\Lambda}_8 \overline{K}_c \Xi
\nonumber \\ && \mbox{}
+ g_{\Lambda_8^{} NK} \overline{\Lambda}_8 \overline{K} N
+ g_{\Xi_8^{} \Xi \pi}^{} \overline{\Xi}_8 \bm{\tau} \cdot \bm{\pi}\, \Xi
\nonumber \\ && \mbox{}
+ g_{\Xi_8^{} \Xi\eta_8^{}}^{} \eta_8 \overline{\Xi}_8 \Xi
+ g_{\Xi_8^{} \Sigma K}^{} \overline{\Xi}_8 \bm{\tau} \cdot \bm{\Sigma} K_c
\nonumber \\ && \mbox{}
+ g_{\Xi_8^{} \Lambda K}^{} \overline{\Xi}_8 K_c \Lambda
+ \mbox{(H.c.)},
\end{eqnarray}
where
\begin{eqnarray}
&& \textstyle
g_{N_8^{}N\pi}^{} = \frac{1}{\sqrt2} (d+f) g_8^{}, \quad
g_{N_8^{}N\eta_8^{}}^{} = - \frac{1}{\sqrt6} (d-3f) g_8^{},
\nonumber \\ && \textstyle
g_{N_8^{}\Lambda K}^{} = - \frac{1}{\sqrt6} (d+3f) g_8^{}, \quad
g_{N_8^{}\Sigma K}^{} = \frac{1}{\sqrt2} (d-f) g_8^{},
\nonumber \\ && \textstyle
g_{\Sigma_8^{} \Sigma \pi}^{} = -\sqrt2 f g_8^{}, \quad
g_{\Sigma_8^{} \Sigma \eta_8^{}}^{} = \sqrt{\frac23} d g_8^{},
\nonumber \\ && \textstyle
g_{\Sigma_8^{} \Lambda \pi}^{} = \sqrt{\frac23} d g_8^{}, \quad
g_{\Sigma_8^{} \Xi K} = - \frac{1}{\sqrt2} (d+f) g_8^{},
\nonumber \\ && \textstyle
g_{\Sigma_8^{} N K}^{} = \frac{1}{\sqrt2} (d-f) g_8^{}, \quad
g_{\Lambda_8^{} \Sigma \pi}^{} = \sqrt{\frac23} d g_8^{},
\nonumber \\ && \textstyle
g_{\Lambda_8^{} \Lambda \eta_8^{}}^{} = - \sqrt{\frac23} d g_8^{}, \quad
g_{\Lambda_8^{} \Xi K}^{} = - \frac{1}{\sqrt6} (d-3f) g_8^{},
\nonumber \\ && \textstyle
g_{\Lambda_8^{} NK}^{} = - \frac{1}{\sqrt6} (d+3f) g_8^{}, \quad
g_{\Xi_8^{} \Xi\pi}^{} = - \frac{1}{\sqrt2} (d-f) g_8^{},
\nonumber \\ && \textstyle
g_{\Xi_8^{}\Xi\eta_8^{}}^{} = -\frac{1}{\sqrt6} (d+3f) g_8^{}, \quad
g_{\Xi_8^{} \Sigma K}^{} = -\frac{1}{\sqrt2} (d+f) g_8^{},
\nonumber \\ && \textstyle
g_{\Xi_8^{} \Lambda K}^{} = - \frac{1}{\sqrt6} (d-3f) g_8^{}.
\end{eqnarray}
With $f=1/2$ and $d=3/2$, the couplings are given explicitly in each channel
in Table~\ref{table-8}.
It can be easily found that the constraint $f/d=1/3$ gives rise to some
selection rules so that the $N_8^{} N \eta_8^{}$, $\Lambda_8 \Xi K$ and
$\Xi_8 \Lambda K$ couplings vanish, as has been noted by Close and Dudek
\cite{CD04} from analyzing the decay of $\Xi$.
Since $\Lambda_8$ and $\Xi_8$ do not mix with the antidecuplet members
due to isospin, the above selection rules hold regardless of the mixing
with the antidecuplet.

\begin{table}[t]
\centering
\begin{ruledtabular}
\begin{tabular}{cc|cc|cc|cc}
\multicolumn{2}{c}{$\Xi_8^-$} & \multicolumn{2}{c}{$\Xi_{8}^0$} &
\multicolumn{2}{c}{$N_{8}^+$} & \multicolumn{2}{c}{$N_{8}^0$}
\\ \hline
$\pi^- \Xi^0$ & $-1$ & $\pi^+ \Xi^-$ & $-1$ & $\pi^0 p$ &
$\sqrt{2}$ &
$\pi^0 n $ & $-\sqrt{2}$ \\
$\pi^0 \Xi^-$ & $\frac{1}{\sqrt2}$ & $\pi^0 \Xi^0$ &
$-\frac{1}{\sqrt{2}}$ & $\pi^+ n$ & $2$
& $\pi^- p$ & $2$ \\
$\eta_8^{} \Xi^-$ & $-\sqrt{\frac{3}{2}}$ & $\eta_8^{} \Xi^0$ &
$-\sqrt{\frac{3}{2}}$ & $\eta_8^{} p$ & $0$ &
$\eta_8^{} n$ & $0$ \\
$\bar{K}^0 \Sigma^-$ & $-2$ & $K^- \Sigma^+$ & $2$ & $K^+
\Sigma^0$ &
$\frac{1}{\sqrt2}$ & $K^0 \Sigma^0$ & $-\frac{1}{\sqrt{2}}$ \\
$K^- \Sigma^0$ & $-\sqrt2$ & $\bar{K}^0 \Sigma^0$ & $-\sqrt2$ &
$K^0 \Sigma^+$ &
$1$ & $K^+ \Sigma^-$ & $1$ \\
$K^- \Lambda$ & 0 & $\bar{K}^0 \Lambda$ & $0$ & $K^+ \Lambda$ &
$-\sqrt{\frac{3}{2}}$ & $K^0 \Lambda $ & $-\sqrt{\frac{3}{2}}$ \\
\hline \multicolumn{2}{c}{$\Sigma_{8}^0$} &
\multicolumn{2}{c}{$\Sigma_{8}^+$} &
\multicolumn{2}{c}{$\Sigma^-_{8}$} &
\multicolumn{2}{c}{$\Lambda^0_{8}$}
\\ \hline
$\pi^+ \Sigma^-$ & $\frac{1}{\sqrt2}$ & $\pi^+ \Sigma^0$ &
$-\frac{1}{\sqrt2}$ & $\pi^-\Sigma^0$ & $\frac{1}{\sqrt{2}}$ & $\pi^- \Sigma^+$ & $\sqrt{\frac{3}{2}}$ \\
$\pi^- \Sigma^+$ & $-\frac{1}{\sqrt{2}}$ & $\pi^0 \Sigma^+$ &
$\frac{1}{\sqrt{2}}$
& $\pi^0 \Sigma^-$ & $-\frac{1}{\sqrt2}$ & $\pi^+ \Sigma^-$ & $\sqrt{\frac{3}{2}}$ \\
$\pi^0 \Lambda$ & $\sqrt{\frac{3}{2}}$ & $\eta_8^{} \Sigma^+$ &
$\sqrt{\frac{3}{2}}$ & $\eta_8^{} \Sigma^-$ & $\sqrt{\frac{3}{2}}$ & $\pi^0 \Sigma^0$ & $\sqrt{\frac{3}{2}}$ \\
$\eta_8^{}\Sigma^0$ & $\sqrt{\frac{3}{2}}$ & $\pi^+ \Lambda$ &
$\sqrt{\frac{3}{2}}$ & $\pi^- \Lambda$
& $\sqrt{\frac{3}{2}}$ & $\eta_8^{} \Lambda$ & $-\sqrt{\frac{3}{2}}$ \\
$K^- p$ & $\frac{1}{\sqrt2}$ & $K^+ \Xi^0$ & $2$ & $K^0 \Xi^-$ & $-2$ & $ K^+ \Xi^-$ & 0 \\
$\bar{K}^0 n$ & $-\frac{1}{\sqrt2}$ & $\bar{K}^0 p$ & 1 & $K^- n$ &1 & $K^0 \Xi^0$ & 0 \\
$K^+ \Xi^-$ & $-\sqrt2$ & & & & & $K^- p$ & $-\sqrt{\frac{3}{2}}$ \\
${K}^0 \Xi^0$ & $-\sqrt2$ & & & & & $\bar{K}^0 n$ &
$-\sqrt{\frac{3}{2}}$ \\
\end{tabular}
\end{ruledtabular}
\caption{Couplings of the pentaquark octet with the baryon octet
and pseudoscalar meson octet.
Multiplying the universal coupling constant $g_8^{}$ is understood.
Terms with 0 mean vanishing couplings, which implies the selection rules.}
\label{table-8}
\end{table}

\subsection{Selection rule in ideal mixing}

When there is ideal mixing between the pentaquark octet and antidecuplet
states for the nucleon and Sigma resonances, the decay will separate into
strange and nonstrange parts.
Within ideal mixing, the nucleons with no $\bar{s}s$ component will decay
into nonstrange hadrons and that with $\bar{s}s$ component will
decay into strange hadrons only.%
\footnote{The nucleons with no $\bar ss$ component may couple to
strange hadrons through quark-antiquark creation.
However, this is possible only when the normal baryon has
five quark component, which is suppressed in the OZI limit.}

To separate the ${\bar s} s$ and light ${\bar q} q$ components in
$\eta_8^{}$, we introduce the interaction of the pentaquark octet with
the normal baryon octet and meson singlet $\eta_1^{}$ given by
\begin{eqnarray}
\mathcal{L}_1 &=& -\sqrt{3} g_1^{} \overline{P}_i^j B^i_j \eta_1^{}
+(\rm H.c.)
\nonumber \\
&=& -\sqrt{3} g_1^{} \eta_1^{} \left( \overline{N}_8 N +
\overline{\Lambda}_8 \Lambda + \overline{\Xi}_8 \Xi +
\overline{\bm{\Sigma}}_8 \cdot \bm{\Sigma} \right) + \mbox{(H.c.)}.
\nonumber \\
\end{eqnarray}
In ideal mixing, the couplings $g_8^{}=g_{\overline{10}}^{}=g_1$
($= g$) will be all equal \cite{CD04} and we have the interactions
as
\begin{eqnarray}
\mathcal{L}_{\textrm{\scriptsize mixing}} &=&
- \sqrt3 g \overline{N}_{\bar{q}q} \bm{\tau} \cdot \bm{\pi} N
\nonumber \\ && \mbox{}
+ \sqrt3 g \left( \eta_{\bar q q} \overline{N}_{\bar qq} - \eta_{\bar ss}
\overline{N}_{\bar ss} \right) N
\nonumber \\ && \mbox{}
- \frac{3}{\sqrt2} g \overline{N}_{\bar ss} K \Lambda
+ \sqrt{\frac32} g \overline{N}_{\bar ss} \bm{\tau} \cdot \bm{\Sigma} K
\nonumber \\ && \mbox{}
+ i \sqrt{\frac32} g \left( \overline{\bm{\Sigma}}_{\bar qq} \times
\bm{\Sigma} \right) \cdot \bm{\pi}
+ \sqrt{\frac23} g \, \eta_{\bar qq} \overline{\bm{\Sigma}}_{\bar qq} \cdot
\bm{\Sigma}
\nonumber \\ && \mbox{}
- \sqrt6 g \,\eta_{\bar ss} \overline{\bm{\Sigma}}_{\bar ss} \cdot
\bm{\Sigma}
- \sqrt3 g \overline{K}_c \overline{\bm{\Sigma}}_{\bar ss} \cdot \bm{\tau} \Xi
\nonumber \\ && \mbox{}
- \frac{3}{\sqrt2} g \overline{\bm{\Sigma}}_{\bar qq} \cdot \bm{\pi} \Lambda
- \sqrt{\frac32} g \overline{K} \, \overline{\bm{\Sigma}}_{\bar qq} \cdot
\bm{\tau} N
\nonumber \\ &&
\mbox{} + \mbox{(H.c.)},
\label{Lag-mix}
\end{eqnarray}
which is obtained by using
\begin{eqnarray}
&& N_{\overline{10}} = \sqrt{\frac13} N_{\bar qq} + \sqrt{\frac23} N_{\bar
ss}, \quad
N_{8} = -\sqrt{\frac23} N_{\bar qq} + \sqrt{\frac13} N_{\bar ss},
\nonumber \\ &&
\Sigma_{\overline{10}} = \sqrt{\frac23} \Sigma_{\bar qq} + \sqrt{\frac13}
\Sigma_{\bar ss}, \quad
\Sigma_{8} = -\sqrt{\frac13} \Sigma_{\bar qq} + \sqrt{\frac23}
\Sigma_{\bar ss}.
\nonumber \\
\end{eqnarray}
One can then look for the pentaquark states which have a hypercharge $1$
and $0$ by looking at the specific decay channels given in
Table~\ref{tab-mix}.

\begin{table}
\centering
\begin{ruledtabular}
\begin{tabular}{cc|cc|cc|cc}
\multicolumn{2}{c}{$N^+_{\bar{q}q}$} &
\multicolumn{2}{c}{$N^+_{\bar{s}s}$} &
\multicolumn{2}{c}{$N^0_{\bar{q}q}$} &
\multicolumn{2}{c}{$N^0_{\bar{s}s}$}
\\ \hline
$\pi^+ n$ & $-\sqrt6$ & $K^+ \Lambda$ & $-\frac{3}{\sqrt2}$ &
$\pi^0 n$ & $\sqrt{3}$ &
$K^0 \Sigma^0 $ & $-\sqrt{\frac{3}{2}}$ \\
$\pi^0 p$ & $-\sqrt3$ & $K^+ \Sigma^0$ & $\sqrt{\frac{3}{2}}$ &
$\pi^- p$ & $-\sqrt6$
& $K^+ \Sigma^-$ & $\sqrt3$ \\
$\eta_{\bar{q}q}^{} p$ & $\sqrt3$ & $K^0 \Sigma^+$ & $\sqrt{3}$ &
$\eta_{\bar{q}q}^{}$ & $\sqrt3$&
$K^0 \Lambda$ & $-\frac{3}{\sqrt2}$ \\
& & $\eta_{\bar{s}s}^{} p$ & $-\sqrt3$ && & $\eta_{\bar{s}s}^{} n$ &
$-\sqrt3$ \\
\hline \multicolumn{2}{c}{$\Sigma_{\bar{q}q}^+$} &
\multicolumn{2}{c}{$\Sigma_{\bar{s}s}^+$} &
\multicolumn{2}{c}{$\Sigma^-_{\bar{q}q}$} &
\multicolumn{2}{c}{$\Sigma^-_{\bar{s}s}$}
\\ \hline
$\pi^+ \Sigma^0$ & $\sqrt{\frac{3}{2}}$ & $K^+ \Xi^0$ &
$\sqrt6$ & $\pi^-\Sigma^0$ & $-\sqrt{\frac{3}{2}}$ & $K^0 \Xi^-$ & $-\sqrt6$ \\
$\pi^0 \Sigma^+$ & $-\sqrt{\frac{3}{2}}$ & $\eta_{\bar{s}s}^{}
\Sigma^+$ & $-\sqrt6$
& $\pi^0 \Sigma^-$ & $\sqrt{\frac{3}{2}}$ & $\eta_{\bar{s}s}^{} \Sigma^-$
& $-\sqrt6$ \\
$\eta_{\bar{q}q}^{} \Sigma^+ $ & $\sqrt{\frac{3}{2}}$ & &
& $\eta_{\bar{q}q}^{} \Sigma^-$ & $\sqrt{\frac{3}{2}}$ & & \\
$\pi^+ \Lambda$ & $-\frac{3}{\sqrt{2}}$ & & & $\pi^- \Lambda$
& $-\frac{3}{\sqrt{2}}$ & & \\
$\bar{K}^0 p$ & $-\sqrt3$ & & & $K^- n$ & $-\sqrt3$ & & \\
\hline \multicolumn{2}{c}{$\Sigma_{\bar{q}q}^0$} &
\multicolumn{2}{c}{$\Sigma_{\bar{s}s}^0$} & \multicolumn{2}{c}{} &
\multicolumn{2}{c}{}
\\ \hline
$\pi^+ \Sigma^-$ & $-\sqrt{\frac{3}{2}}$ & $K^+ \Xi^-$ &
$-\sqrt3$ & & & \\
$\pi^- \Sigma^+$ & $\sqrt{\frac{3}{2}}$ & $K^0 \Xi^0$ & $-\sqrt3$
& & & & \\
$\eta_{\bar{q}q}^{} \Sigma^0 $ & $\sqrt{\frac{3}{2}}$
& $\eta_{\bar{s}s}^{} \Sigma^0$ & $-\sqrt6$ & & & & \\
$\pi^0 \Lambda$ & $-\frac{3}{\sqrt{2}}$ & & & & & & \\
$K^- p$ & $-\sqrt{\frac{3}{2}}$ & & & & & \\
$\bar{K}^0 n$ & $\sqrt{\frac{3}{2}}$ & & & & & \\
\end{tabular}
\end{ruledtabular}
\caption{Couplings of the ideally mixed pentaquark baryon states.
The subscripts represents states with either pure light
quark-antiquark pair or pure strange quark-antiquark pair.
Multiplying the universal coupling constant $g$ is understood.}
\label{tab-mix}
\end{table}

\section{Coupling with baryon decuplet}

We now consider the interactions of pentaquark baryons with baryon decuplet
and meson octet.
As shown in Ref.~\cite{OKL03b}, the pentaquark antidecuplet cannot couple
to the baryon decuplet and meson octet.
However, the pentaquark octet may couple to the decuplet-octet channel,
which can lead to a nonvanishing coupling for $N_{8} \Delta \pi$.
But in this case we can not have a separation as in Eq.~(\ref{Lag-mix})
since $N_{\overline{10}} \Delta \pi$ coupling is now not allowed.
This means that there can exist $N_{\bar ss} \Delta\pi$ coupling, which,
however, violates the OZI rule \cite{CD04}.
Therefore, the OZI rule prohibits the coupling of the pentaquark octet with
the baryon decuplet and meson octet unless other OZI-evading processes
are allowed.

With our rule for tensor notation, this can be seen easily.
That is, we can have a coupling between the pentaquark octet and baryon
decuplet and meson octet in the form of
\begin{eqnarray}
\mathcal{L}_{\rm int} &=& h_{8}^{} \,
\epsilon^{ijk}\epsilon^{lpq} \overline{S}_{[ip,q]} D_{jlm} M_k^m +
\mbox{(H.c.)}
\nonumber \\
&=& 3 \, h_{8}^{} \, \epsilon^{ijk} \overline{P}^l_i D_{jlm} M^m_k +
\mbox{(H.c.)}.
\label{interaction8}
\end{eqnarray}
However, as one can see, the antiquark index $q$ in $S_{[ip,q]}$
is not contracted with the index $m$ in the meson field $M$.
Hence, unlike Eq.~(\ref{interaction2}), the above equation does
not respect the OZI rule.
In this case, one can not make any SU(3) invariant terms where the index
$q$ in the pentaquark field is contracted with the index $m$ in the meson
field.
This implies that $h_{8}^{} = 0$ in the OZI limit, and, therefore, rules
out $N(1440)$ and $N(1710)$ as ideally mixed states of pentaquark octet and
antidecuplet because the branching ratios of their decays into $\Delta\pi$
are large.
These decays are not possible even when the SU(3) symmetry breaking term is
allowed for the interaction of pentaquark octet with baryon decuplet and
meson octet.
A leading order term that respects the generalized OZI rule can be
written as
\begin{eqnarray}
\mathcal{L}_{\rm int} &=& h'_{8} \, \epsilon^{ijk}\epsilon^{lpt}
\bar{S}_{[ip,q]} D_{jlm} M_k^q Y_t^m + \mbox{(H.c.)},
\label{interaction9}
\end{eqnarray}
where $Y$ is the hypercharge operator.
Since this symmetry breaking term still stisfies isospin symmetry,
the interaction lagrangian is not affected in the SU(2) sector, i.e.,
$N_8 \Delta\pi$.
Therefore, the decay $N_8 \to \Delta \pi$ is not allowed even in
the SU(3) symmetry breaking interaction (\ref{interaction9}).
One can explicitly work out the contractions to find that the decay is
indeed forbidden.
This shows that the $N_8 \to \Delta \pi$ decay is an isospin-symmetry
breaking process, i.e., it is allowed only when we include the SU(2)
symmetry breaking terms.

\section{Closing}

We have discussed how the generalized OZI rule leads to ideal mixing and
selection rules in the pentaquark decays.
We have then introduced a simple tensor method that gives rise to the
selection rules in the pentaquark baryon decays.
Through experimental verification of the possible decay channels and
their relative strength, one will be able to identify the other possibly
narrow pentaquark states and test the picture of ideal mixing of pentaquark
octet and antidecuplet states.

\section{Acknowledgments}

SHL would like to thank Tom Cohen and Ismail Zahed for useful discussions.
This work was supported by the Brain Korea 21 project of Korean Ministry
of Education and by KOSEF under Grant No.  1999-2-111-005-5.

\end{document}